\documentclass[preprint,12pt]{elsarticle}

\begin{document}
\begin{frontmatter}
\title{Exact three spin correlation function relations for the square and the honeycomb Ising lattices}
\author[mymainaddress]{T. Kaya}
\ead{tkaya@yildiz.edu.tr}
\address[mymainaddress]{ Physics Department, Y{\i}ld{\i}z Technical University,
        34220 Davutpa\c sa, Istanbul, Turkey}

\begin{abstract}
In this work, the order parameter and the two-site correlation functions are expressed properly using the decimation transformation process in the presence of an external field so that their applications lead to some significant physical results. Indeed, their applications produce or reproduce some relevant and important results which were included  in cumbersome mathematics in the previous studies,  if not in a form impossible to understand. The average magnetization or the order parameter $<\!\!\sigma\!\!> $ is expressed as $<\!\!\sigma_{0,i}\!\!>=
 <\!\!\tanh[ \kappa(\sigma_{1,i}+\sigma_{2,i}+\dots +\sigma_{z,i})+H]\!\!> $. Here, $\kappa$ is the coupling strength, $z$
is the number of nearest neighbors.  $\sigma_{0,i}$ denotes the
central spin at the $i^{th}$ site, while $\sigma_{l,i}$,
$l=1,2,\dots,z$ are the nearest neighbor spins around the central
spin. $H$ is the normalized external magnetic field. We show that the application of this relation to the 1D
Ising model reproduces readily the previously obtained exact
results in the absence of an external field. Furthermore, the
three-site correlation functions of square and honeycomb lattices
of the form $<\!\!\sigma_{1}\sigma_{2}\sigma_{3}\!\!>$  are
analytically obtained. One finds that the three-site correlation
functions are equal to $f(\kappa)\!\!<\!\!\sigma\!\!>$. Here
$f(\kappa)$ depends on the lattice types and is an analytic
function of coupling constant. This result indicates that the
critical properties of three-site correlation functions of those
lattices are the same as the corresponding order parameters
$<\!\!\sigma\!\!>$ of those lattices.  This will mean that the uniqueness of the average magnetization as an order parameter is questionable. 
The application of the two site correlation function relation obtained in this paper produces a relevant exact relation between four site correlation function and two site correlation functions for square lattice. It also produces a relation between correlation functions of honeycomb lattice. Making use of this relation leads to the exact calculation of the constant $A$ appearing in the proposed correlation function relation in the scaling theory. It is also indicated that the correlation length $\xi(\kappa)$ can be obtained exactly in the realm of scaling theory from the obtained correlation function relation for the honeycomb lattice.
In addition, the average magnetization relation obtained in this paper leads easily to the result of the conventional
mean field critical coupling strength which, as a first
approximation, is equal to $\kappa_{c}=1/z$. It is also shown that
systematic improvements of the values of the critical coupling
strengths are possible with mean field treatments in this picture.
\end{abstract}
\begin{keyword}
 Ising model \sep Phase transition \sep Decimation transformation \sep Order parameter
\end{keyword}
\end{frontmatter}

\section{Introduction}
In recent decades, there has been intensive research on phase transitions and critical phenomena \cite{Stanley,Hu,Zhu}. A popular model for such studies is the Ising model \cite{Stanley,Hu,Ising}. The Ising model has been used to study  many physical systems \cite{Sarli,Keskin, Aouini}. In one-dimension (1D), the model
has an exact solution with the prediction of the absence of phase
transition at finite temperatures \cite{Ising, Huang, Kramers,
Baxter}. The Ising model in 2D and 3D lattices has also been
subject of extensive studies. It is well-known that there are no
exact solutions for 3D lattice, whereas exact treatments of 2D
lattices are available \cite{Onsager,Yang,Wu,Sh}. Exactly solvable
lattices in 2D, therefore, have special statuses for investigating
order-disorder phase transitions. They are important not only for
explicitly demonstrating the existence of phase transition at
finite temperatures, but also because they established a benchmark
for simulations, motivated the research for exact solutions of
other models and have served as a testing ground for the
efficiency of new theoretical ideas.

As mentioned above, after cumbersome transfer matrix calculations,
some exact expressions  are available for 2D lattices, especially
for the square lattice. It may be, therefore, important to obtain
new exact expressions with simple mathematics, from the known
exact results. Indeed, that is the main purpose of this paper,
since extra exact expressions may be important for tackling
unsolved problems connected with the Ising model in general. To do so,
I think that the method of decimation transformation, in the
presence of external field, can be useful. At this point, it is
important to express that the method is the starting point of the
well-known real space renormalization group (RSRG) theory
\cite{Wilson, Kadanof, Kadanof1, Swedsen, Tuncer}. The decimation
transformation of the 1D Ising model left the form of the
corresponding Hamiltonian invariant, introducing no new
interactions: the nearest-neighbor are retained, but with a
renormalized interaction. For the 1D Ising
chain, therefore, this method reproduces the result of the exact solution of transfer
matrix method. Exact renormalization group transformations for positive symmetric matrices and one-dimensional q-state Potts model had been constructed \cite{Hu1,Hu2}. However, in general, decimation transformation for
2D and 3D lattices  in RSRG theory produce a hierarchy of
interactions that are consistent with the original lattices. For
this method to be practical, the lower terms must  dominate the
energy, so that the additional terms can be treated
perturbatively. The decimation transformation and RSRG calculations for the two-dimensional Ising model had been formulated \cite{Hu3,Hu4}.  In other words, the exact treatments from  RSRG
method are impossible except for the 1D Ising chain. Therefore, some
other procedure is necessary to obtain new exact relations
\cite{Yalabk,Hilhorst,Atanas,Creutz, Melson}.

To this end, in the following section, we are going to show that
the order parameter and two-site nearest-neighbor correlation
function for Ising lattices can be expressed as
\begin{eqnarray}
&&\!\!\!\!\!\!\!\!\!\!\!\!<\!\!\sigma_{0,i}\!\!>=<\!\!\tanh[\kappa(\sigma_{1,i}+\sigma_{2,i}+\dots+\sigma_{z,i})+H]\!\!>{}\nonumber\\&&
\!\!\!\!\!\!\!\!\!\!\!\!<\!\!\sigma_{0,i}\sigma_{1,i}\!\!>=\frac{1}{z}<\!\!(\sigma_{1,i}+\dots+\sigma_{z,i})\tanh[\kappa(\sigma_{1,i}
+\sigma_{2,i}+\dots+\sigma_{z,i})\!\!+\!\!H]\!\!>\nonumber
\end{eqnarray}
with the help of the decimation transformation in the presence of
external field. These two relations are going to be the corner
stone of the investigation of this paper. The reliability of an
approach may be assessed by seeing whether it reproduces already
evaluated quantities. To see the reliability and feasibility of
these relation at $H=0$,  one can consider the first order mean
field treatment of the order parameter relation as an example.
This is going to be taken into account in the following sections.
Besides, the improvement of the estimation of critical coupling
strength is going to be presented in a tractable manner. It is
also important to investigate the 1D Ising chain to assess the
quality of the method used in this paper. Indeed, it is going to
be shown, in the next section, that the exact treatment of the 1D
chain reproduces exactly the same result of previously obtained
analytical relation.

In addition, in next section, we are going to execute some exact
calculations for the 2D  square and honeycomb Ising lattices using a
more tractable and  simpler mathematics. This is, indeed, our
main focus and motivation of writing this paper. Furthermore, no
new physical assumption is  going to be introduced, except
using the definition of the partition function and the definition
of the order parameter.

This paper is organized as follows. In the next section, the order
parameter and the nearest neighbor two-site correlation function
relations are going to be derived with help of the decimation
transformation in the presence of an external field.  In the third
section, some exact calculations of the square Ising lattice are
going to be presented. In addition, some mean field considerations
for this lattice is going to be pointed out. In the last section,
the honeycomb Ising lattice is going to be considered in the same
manner as in the case of square lattice. Some short discussion
on this paper is going to be also given in the same section.

\section{Decimation transformation in the presence of external field}

The fundamental idea of  the decimation transformation is to
calculate the partition function of a system by successively
thinning out its degrees of freedom. Upon such a partial trace
operation, a new system appears, equivalent to the original one, but with fewer degrees of freedom. To achieve this procedure, the Hamiltonians of the Ising systems need to be written in a  proper form. Assuming, there are $N$ sites in a lattice, writing the Hamiltonian, in the
 presence of external field, in the following form,
\begin{equation} 
\mathcal{H}\!=\!\!-J\sum_{i=1}^{N/2} \sigma_{0,i}(\sigma_{1,i}+\sigma_{2,i}+\dots+\sigma_{z,i})-h\sum_{i=1}^{N/2}[ \sigma_{0,i}+\frac{1}{z}(\sigma_{1,i}+\sigma_{2,i}+\dots+\sigma_{z,i})],
\end{equation}
is very useful for the decimation transformation. Here,
$\sigma_{0,i}$ is the central spin in the $i^{th}$  site (or
block) and $\sigma_{1,i}$ , $\sigma_{2,i}$ and $\dots
\sigma_{z,i}$  are the nearest neighbor spins around the central
spin $\sigma_{0,i}$. $J$ is the interaction constant, $z$ is the
number of nearest neighbors, and $H$ is the external magnetic
field. The partition function for this Hamiltonian can be written
as
\begin{eqnarray}
 Z\!=\!\sum_{\{\sigma\}}\!\! e^{\textstyle \kappa\! \sum \sigma_{0,i}(\sigma_{1,i}\!+\!
\sigma_{2,i}+\dots+\sigma_{z,i})
\!+\!H\sum[\sigma_{0,i}\!+\!\frac{1}{z}(\sigma_{1,i}\!+\!\sigma_{2,i}\!+\!\dots\!+\!\sigma_{z,i})]}.
\end{eqnarray}
Here $\kappa=\frac{J}{kT}$ is the coupling constant, $k$ is
Boltzmann constant, and $T$ is the temperature of the system.
$H=\frac{h}{kT}$ is the normalized external magnetic field which
can be called simply external field from now on. Apparently, the
set $\{\sigma\}$ is equal to the union of $\{\sigma_{0,i}\}$  and
$\{\sigma_{l,i}\}$. Taking into account the sets
$\{\sigma_{0,i}\}$ and $\{\sigma_{\sigma_{l,i}}\}$, here $l$ runs
from $1$ to $z$, independently, the partition function can be
rearranged as
\begin{equation}
Z\!=\!\!\!\sum_{\{\sigma_{l,i}\}} e^{\ H\sum\frac{1}{z}(\sigma_{1,i}+
\sigma_{2,i}+\dots+\sigma_{z,i})}\!\prod_{\{\sigma_{0,i}\}}\!e^{\kappa \sum \sigma_{0,i}(\sigma_{1,i}+
\sigma_{2,i}+\dots+\sigma_{z,i})+H\sum \sigma_{0,i}}.
\end{equation}
Since there are no interactions among spins in the set
$\{\sigma_{0,i}\} $, the terms in the product can be evaluated
quite easily by substituting $\pm 1$ values to the set variable
$\sigma_{0,i}$ and the partition function turns out to be
\begin{equation}
Z\!=\!\!\!\sum_{\{\sigma_{l,i}\}}e^{ H\sum\frac{1}{z}
(\sigma_{1,i}\!+\!\sigma_{2,i}\!+\!\dots\!+\!\sigma_{z,i})}\!\!\! \prod_{\{\sigma_{0,i}\}}\!\!2\cosh[\kappa (\sigma_{1,i}\!+\!\sigma_{2,i}\!+\!\dots\!+\!\sigma_{z,i})+H].
\end{equation}
Rewriting the product in the form of summation, the partition
function turns out to be
\begin{equation}
Z\!=\!\!\!\sum_{\{\sigma_{l,i}\}}\!\!e^{ H\sum\frac{1}{z}
(\sigma_{1,i}+\sigma_{2,i}+\dots+\sigma_{z,i}) +\sum\ln[2\cosh(\kappa (\sigma_{1,i}+\sigma_{2,i}+\dots+\sigma_{z,i})\!+\!H]}.
\end{equation}
Apparently, the partition functions presented in Eq. (2) and Eq. (5)
are equivalent. This is the only physical significance of making
the above decimation transformation. Now it is time to put this
equivalence into practice. Taking the natural logarithm of
partition functions in Eq. (2) and Eq. (5) respectively and
differentiating both with respect to $H$  leads readily to
\begin{equation}
<\!\!\sigma_{0,i}\!\!>=<\!\!\tanh[\kappa(\sigma_{1,i}+\sigma_{2,i}+\dots+\sigma_{z,i})+H]\!\!>.
\end{equation}
Doing the same procedure to the partitions functions and differentiating with respect to $\kappa$ easily produce the following relation,
\begin{equation}
<\!\!\sigma_{0,i}\sigma_{l,i}\!\!>=\frac{1}{z} <(\sigma_{1,i}+\sigma_{2,i}+\dots+\sigma_{z,i})\tanh[\kappa
(\sigma_{1,i}+\sigma_{2,i}+\dots+\sigma_{z,i})+H]\!\!>.
\end{equation}
Before starting the applications of Eq. (6) and Eq. (7), it is important to point out that there are  similar papers in literature \cite{y1,y2}, but they do not carry the intentions which we are going to consider in this paper. 

Let us  now  consider some simple applications of these two
relations to appreciate the relevance and significance of our
procedure. The reliability of an approach may be assessed by
seeing whether it reproduces already evaluated quantities. To see
the reliability and feasibility of these relations at $H=0$,  one
can consider the first order mean field treatment of the order
parameter relation as an example. Applying first order
approximation to the order parameter relation, it can be written
as $<\!\!\sigma\!\!>=\tanh(z\kappa<\!\!\sigma\!\!>)$. It is known
from phase transition theory that  $<\!\!\sigma\!\!>=0$, if
$\kappa$  is less than the critical value $\kappa_{c}$,  and
$<\!\! \sigma\!\!>$ differs from zero only  if $\kappa$ is greater than the
critical value. Thus, around the critical point, it can be written
as $<\!\!\sigma\!\!>=z\kappa<\!\!\sigma\!\!>$. In general, this final
relation, the first order mean field critical coupling values of the Ising lattices,  can be expressed as $\kappa_{c}=\frac{1}{z}$.
It is important to notice that the mathematical as well as physical consideration in the derivation of the first order critical coupling values are quite relevant and tractable. On the other hand, to justify the physical relevancy and meaning of the first order relation, the authors of the text books are separated pages. This is same even in a recently published book \cite{book}. Therefore, the simplicity and relevancy of the derivation of the first order relation is significant and important from both mathematical and physical points of view.  

Another example to see the feasibility of order parameter
relation, one can consider the 1D Ising chain. In this case the
order parameter relation can be expressed as
$<\!\!\sigma_{0}\!\!>=<\!\!\tanh\kappa(\sigma_{1}+\sigma_{2})\!\!>$.
Considering the set $\{\sigma\}$,  which takes the values $\pm1$,
$\tanh\kappa(\sigma_{1}+\sigma_{2})$ can be equivalently written
as $\frac{1}{2}(\sigma_{1}+\sigma_{2})\tanh(2\kappa)$. Since the
average value of $<\!\!\sigma_{0}\!\!>$,  $<\!\!\sigma_{1}\!\!>$,
and $<\!\!\sigma_{2}\!\!>$ are equal, let us use  $<\!\!\sigma\!\!>$.
Then this relation can be written as
$<\!\!\sigma\!\!>=<\!\!\sigma\!\!>\tanh(2\kappa)$. Considering the
phase transition concepts, it is easy to see from this final
relation that $<\!\!\sigma\!\!>$ is zero for any finite value of
$\kappa$ and it can be different from zero if only $ \kappa$  goes
to infinity. At this point, it is important to express that the
application of the order parameter relation to the 1D Ising chain
reproduces readily the result of exact solutions which have been
obtained by other methods. The corresponding calculation for the
2D Ising lattices, namely square and honeycomb, are going to be
the subject of section $3$ and section $4$ respectively.

\section{The square lattice Ising model}
In this section, the application of the obtained relation is going to be presented with a more representative manner.
To do so, we need to first write the above relations in the case of square lattice. Considering that square lattice has four nearest neighbors,
these relations arranged for zero external field case as follows
\begin{equation}
<\!\!\sigma\!\!>=<\!\!\tanh[\kappa(\sigma_1+\sigma_2+\sigma_3+\sigma_4)]\!\!>,
\end{equation}
\begin{equation}
<\!\!\sigma_{0}\sigma_{1}\!\!>=\frac{1}{4} <(\sigma_1+\sigma_2+\sigma_3+\sigma_4)\tanh[\kappa(\sigma_1+\sigma_2+\sigma_3+\sigma_4)]\!\!>.
\end{equation}

It is important to notice that, we ignore the indices $i$  in the last relations for the sake of simplicity, since those
equations are the same for each block. It is also important to notice that the average magnetization of each
central spin $<\!\!\sigma_{0,i}\!\!>$ and the average of each nearest neighbor spins $<\!\!\sigma_{l,i}\!\!>$ are equal to each other.
That is why, all of them are denoted by the notation $<\!\!\sigma\!\!>$.
In addition, all of the correlations between the central spin and one of its nearest neighbor
$\!\!<\sigma_{0,i}\sigma_{l,i}\!\!> $ are equivalent. That is why they all replaced by notation $<\!\!\sigma_{0}\sigma_{1}\!\!>$.
After clarifying our notation, now it is time to go on our calculation.

With these neat and simple relations about the average
magnetization and nearest neighbor two site correlation functions,
one can get some useful equalities relating the order parameter
and correlation functions of the system. In other words, we hope
that these relations may enable us to produce some relevant
interrelations between the order parameter and the correlation
functions of the system. At this point, it is worth  mentioning
that the more nearest neighbor a lattice has, the more correlation
function terms can appear at the end of the calculation. As seen
in our above treatment for the 1D Ising chain, the exact solution
can be evaluated quite easily. This is  because, the 1D chain has
only two nearest neighbors which can not produce very complicated
relations. If we turn back to the treatment of  the 2D Ising
square lattice problem, apparently, the first step in the
treatment is to express the
$\tanh[\kappa(\sigma_1+\sigma_2+\sigma_3+\sigma_4)]$ term more
properly so that the averaging process can be evaluated
relevantly. The following simple equality can be useful for this
purpose. Thus,
\begin{equation}
\tanh[\kappa(\sigma_1+\sigma_2+\sigma_3+\sigma_4)]=\frac{\tanh[\kappa(\sigma_1+
\sigma_2)]+\tanh[\kappa(\sigma_3+\sigma_4)]}{1+\tanh[\kappa(\sigma_1+\sigma_2)]\tanh[\kappa(\sigma_3+\sigma_4)]}.
\end{equation}
The term $\tanh[\kappa(\sigma_1+\sigma_2)]$ can be equivalently expressed as
$ \frac{1}{2}(\sigma_+\sigma_2)\tanh(2\kappa)$. Using this identity, the last equation can be written as
\begin{equation}
\tanh[\kappa(\sigma_1+\sigma_2+\sigma_3+\sigma_4)]=\frac{\frac{1}{2}[(\sigma_1+\sigma_2)+
(\sigma_3+\sigma_4)]\tanh(2\kappa)}{1+\frac{1}{4}(\sigma_1+\sigma_2)(\sigma_3+\sigma_4)\tanh^{2}(2\kappa)}.
\end{equation}
Now, the term in the denominator of last equation can be expressed as
\begin{eqnarray}
&&\textstyle [1+
\frac{1}{4}(\sigma_1+\sigma_2)(\sigma_3+\sigma_4)\tanh^{2}(2\kappa)]^{-1}={}\nonumber\\&&
 \sum_{n=0}^\infty(-1)^{n}[\frac{1}{4}(\sigma_1+\sigma_2)(\sigma_3+\sigma_4)]^{n}\tanh^{2n}(2\kappa).
\end{eqnarray}
Multiplying this relation with the numerator of Eq.(11), and doing some algebra, leads to
\begin {eqnarray}
\!\!\!\!\!\!\!\!\!\!\!\!\!\!\!\!\!\!\!\!\!\!\!\!\!\!\!\!\!\!&&\textstyle\tanh[\kappa(\sigma_1+\sigma_2+
\sigma_3+\sigma_4)]=\frac{1}{2}(\sigma_1+\sigma_2+\sigma_3+\sigma_4)\tanh(2\kappa){}\nonumber\\
&&+\frac{1}{4}(\sigma_1+
\sigma_2+\sigma_3+\sigma_4+\sigma_{1}
\sigma_{2}\sigma_{3}+\sigma_{1}\sigma_{2}\sigma_{4})
\sum_{n=1}^{\infty}(-1)^{n}\tanh^{2n+1}.
\end{eqnarray}
Now, evaluation of the averaging of this last relation is quite straightforward. Thus, the average magnetization relation in Eq.(8)
can be expressed as,
\begin{equation}
<\!\!\sigma\!\!>=2<\!\!\sigma\!\!>[\tanh(2 \kappa)\!-\!\frac{\tanh^{3}2\kappa}{1+\tanh^{2}(2 \kappa)}]
-\!<\!\!\sigma_{1}\sigma_{2}\sigma_{3}\!\!>\frac{\tanh^{3}(2 \kappa)}{1+\tanh^{2}(2 \kappa)},
\end{equation}
where we have used $<\sigma_{1}\sigma_{2}\sigma_{3}>=<\sigma_{1}\sigma_{2}\sigma_{4}>$. If this last equation is simplified, it appears as,
\begin{equation}
<\sigma>=-\frac{\tanh^{3 } (2 \kappa)}{1+\tanh^{2} (2 \kappa)-2 \tanh(2 \kappa)}<\sigma_{1}\sigma_{2}\sigma_{3}>.
\end{equation}
The physical importance of this last relation is that there is an exact expression for the average magnetization
 $<\sigma> $ obtained by Yang, which was given as,
$<\sigma>=0$, if the coupling strength is less than the critical
value $\kappa_{c}$. It has values different from zero for
$\kappa\geq\kappa_{c}$, which was expressed as
$<\sigma>=[1-\sinh^{-4}(\kappa)]^{1/8}$. At this point, it is
worth  mentioning that the exact average magnetization relation
was first derived by Onsager (1949), but he never published the
details of his calculation \cite{Beale}. Using this exact result
for average magnetization, the exact expression for
$<\sigma_{1}\sigma_{2}\sigma_{3}>$  can be written as,
\begin{equation}
<\sigma_{1}\sigma_{2}\sigma_{3}>= \left\{\begin{array}{ll} 0 &\:\:\:\textrm{for}\:\:\:\kappa\leq\kappa_{c}\\
f(\kappa)
[1-\sinh^{4}(\kappa)]^{1/8}&\:\:\:\textrm{for}\:\:\:\kappa\geq\kappa_{c}
\end{array} \right.
\end{equation}
where $f(\kappa)$ is apparently equal to $[1+\tanh^{2} (2
\kappa)-2 \tanh(2 \kappa)]\tanh^{-3}(2\kappa)$. Finally, we have
succeeded in deriving an exact analytical expression for the three
site correlation function of the square Ising lattice of the form
$<\sigma_{1}\sigma_{2}\sigma_{3}>$. Apparently, one can easily see
from the last equation that the critical behavior of three site
correlation function is the same as the average magnetization
which is considered as unique order parameter for the Ising
systems. I find this result very interesting because it indicates
that the tree site correlation function can also be considered as
the order parameter for the square lattice Ising model. Of course,
it is not easy to make further comments about this result which 
can not be expected intuitively.  But it is just a mathematical
fact.

Let us now examine the order parameter relation from the mean field aspect. For a crude approximation, the three site correlation
function can be approximately replaced by $<\sigma>^{3}$.  Substituting this approximate value into Eq. (14)
 leads to the following relation at critical point,
\begin{equation}
2[\tanh(2 \kappa_{c})-\frac{\tanh^{3}2\kappa_{c}}{1+\tanh^{2}(2 \kappa_{c})}]=1.
\end{equation}
From this relation, the value of $\kappa_c$ is estimated to be $0.37$. Now, instead of using this rough approximation,  we note that if three site correlation
function is approximated as $<\sigma><\sigma_2\sigma_3>$, Eq. (14) produces the following relation at critical point,
\begin{equation}
\!\!\!\!\!\!\!\!\!\!\!\!\!\!\!\!\!\!\!\!\!2[\tanh(2 \kappa_{c})-\frac{\tanh^{3}2\kappa_{c}}{1+\tanh^{2}(2 \kappa_{c})}]
-\frac{\tanh^{3}(2 \kappa)_{c}}{1+\tanh^{2}(2 \kappa_{c})}
<\sigma_{2}\sigma_{3}>\!\!|_{\kappa=\kappa_c}=1.
\end{equation}
The values of $<\sigma_{2}\sigma_{3}>\!\!|_{\kappa=\kappa_c}$ are
obtained in references \cite{McCoy,McCoy1} as $0.636$. Thus
substituting this value in the last relation, one can find  the
critical value of coupling strength as $0.43$, which is very close to
the exact value $\kappa_ {c}=0.4407$.  From the last two
considerations to obtain critical coupling value, we can claim that the method used in this paper is a quite relevant approach for the application of the mean field treatment in a tractable manner. If some of the previous mean field treatments \cite{Weiss,Wysin,Kaya1,Oguchi,Van,Linens,Kaya2} are considered, one can easily see the relevancy.
The next step in the investigation of the 2D square lattice Ising model of this section is to calculate a proper expression for
the nearest neighbor two site correlation function given by Eq. (9). After some calculation and averaging, if the relation present in Eq. (13) is multiplied by
$(\sigma_1+\sigma_2+\sigma_3+\sigma_4)$, , Eq. (9) can be expressed as
\begin{eqnarray}
&&\!\!\!\!\!\!\!\!\!\!\!\!\!\!\!\!\!\!\!\!\!\textstyle<\sigma_{0}\sigma_{1}>=\frac{1}{4}\tanh(2\kappa)+\frac{1}{8}\tanh{4\kappa}
+\frac{1}{4}\tanh(4\kappa)[2<\sigma_1\sigma_{3}>+<\sigma_{1}\sigma_{2}>]
{}\nonumber\\
&&\:\:\:+\frac{1}{4}[-\tanh(2\kappa)+\frac{1}{2}\tanh(4\kappa)]<\sigma_1\sigma_2\sigma_3\sigma_4>,
\end{eqnarray}
where $<\!\! \sigma_1\sigma_{2}\!\!>$ is the next nearest neighbor correlation function and $<\!\! \sigma_1\sigma_3\!\!>$ is the second next nearest
neighbor correlation function. The exact result for all of the two site correlation functions were obtained  by McCoy.
This means that the four site correlation function can be
expressed exactly as,
\begin{eqnarray}
&&<\sigma_1\sigma_2\sigma_3\sigma_4>=[\frac{1}{4}[-\tanh(2\kappa)+\frac{1}{2}\tanh(4\kappa)]^{-1}\{<\sigma_{0}\sigma_{1}>
{}\nonumber\\&&-\frac{1}{4}\tanh(2\kappa)-\frac{1}{8}\tanh{4\kappa}-\frac{1}{4}\tanh(4\kappa)[2<\sigma_1\sigma_{3}>+<\sigma_{1}\sigma_{2}>]\}.
\end{eqnarray}
Before we close this section, a few remarks seems to be in order.
The three site correlation function and the four site correlation
function of the 2D square Ising lattice are exactly derived quite
readily. Therefore, one can conclude that the approach of this
paper is exceedingly  simpler than using Pfafian type
calculations. Indeed, there is no exact treatments of three and
four site correlation functions in literature. Besides explicitly
evaluating these correlation functions, a more completely
understanding of their physical interpretations is also important.
The relation obtained for the three site correlation function in
this work forced us to interpret it as a new candidate for the
order parameter of the system besides the average magnetization. I
think that this final remark is the most important physical
consequence of this section as well as analytical calculations of
the correlation function relations. Tractable mean field treatment
of our approach can be also considered as a moderate physical
consequence of this section. Inspiring from these important
results, now it is time to proceed further. Indeed, in the next
section, we are going to study the 2D honeycomb Ising lattice as a
second example of the application of decimation transformation
approach.

\section{Honeycomb lattice}
In this section, we are going to apply the decimation
transformation approach to the honeycomb lattice. Our starting
point is to arrange Eq. (6) and Eq. (7) for the honeycomb lattice
structure. Taking into account its three nearest neighbors and
using these equations' independence from the lattice blocks, which
is denoted by the indices $i$  for each block, those equations can
be expressed as
\begin{equation}
<\!\!\sigma\!\!>=<\!\!\tanh[\kappa(\sigma_1+\sigma_2+\sigma_3]\!\!>,
\end{equation}
\begin{equation}
<\!\!\sigma_{0}\sigma_{1}\!\!>=\frac{1}{3} <(\sigma_1+\sigma_2+\sigma_3)\tanh[\kappa(\sigma_1+\sigma_2+\sigma_3)]\!\!>.
\end{equation}
To facilitate the averaging calculations appearing on the right
sides of these equations, writing an equivalent form of the
function, $\tanh{\kappa(\sigma_1+\sigma_2+\sigma_3)}$, may be
really helpful for our purpose as we have done for the 2D square
lattice case. In this respect, the following equivalent form of
this function may be quite helpful.
\begin{equation}
\tanh[\kappa(\sigma_1+\sigma_2+\sigma_3]=\frac{\frac{1}{2}(\sigma_1+\sigma_2)\tanh(2\kappa(\sigma_1+\sigma_2))+\sigma_{3}\tanh\kappa}
{1+\frac{1}{2}(\sigma_1+\sigma_2)\sigma_3\tanh(2\kappa)\tanh\kappa}.
\end{equation}
Once again,  inverse of the denominator may be expanded into a
series as
\begin{equation}
\frac{1}{1+\frac{1}{2}(\sigma_1+\sigma_2)\sigma_3\tanh(2\kappa)\tanh\kappa}=
\sum_{n=0}^{\infty}(-1)^{n}[\frac{1}{2}(\sigma_1+\sigma_2)\sigma_3\tanh(2\kappa)\tanh\kappa]^{n}.
\end{equation}
Now multiplying this result with the numerator of  Eq. (23) and
taking the averaging calculation into account leads to
\begin{eqnarray}
\!\!\!&&<\!\!\sigma\!\!>=\{\tanh(2\kappa)+\tanh(\kappa)-a(\tanh(2\kappa)+2\tanh\kappa){}\nonumber\\
&&+b(2\tanh(2\kappa)\!\!
+\!\!\tanh\kappa)\}\!\!<\!\!\sigma\!\!>+[b\tanh\kappa-a
\tanh(2\kappa)]\!\!<\!\!\sigma_{1}\sigma_{2}\sigma_{3}\!\!>,
\end{eqnarray}
where $b=a\tanh(2\kappa)\tanh\kappa$, and $a$ satisfies the
following relation
\begin{equation}
a=\frac{\tanh(2\kappa)\tanh\kappa}{2(1-\tanh^{2}(2\kappa)\tanh^{2}\kappa)}.
\end{equation}
Eq. (25) relates the order parameter of an three site correlation function of honeycomb Ising lattice. Unluckily, there is no exact treatment of the order parameter of the honeycomb lattice in literature.
It is, therefore, impossible to obtain an exact expression for three site correlation function of honeycomb lattice, similar to 
what has been obtained in the case of square lattice and expressed with Eq. (16).
For our purpose, the lack of the exact expression for the order parameter is not that important,
since we know from the second order phase transition that the average magnetization is an order parameter regardless of the lattice type.
That means, the values of the order parameter is zero if $\kappa\leq\kappa_c$ whereas it differs from zero if $\kappa\geq\kappa_c$.
Here $\kappa_c$ is the value of the critical coupling strength of the honeycomb lattice. Making use of this knowledge, one can easily claim that
the three site correlation function appearing in Eq. (25) is also zero, if $\kappa\leq\kappa_c$ and it has non-zero values if $\kappa\geq\kappa_c$.
In other words, it behaves as the order parameter of the system. Apparently, we have reached the same conclusion as in the case of square lattice.

Let us now examine Eq. (25) in the mean field manner. For a rough mean field treatment, the three site correlation function in Eq. (25) can be
approximated as $<\sigma>^{3}$. Substituting this approximate value into Eq. (25) and considering the equation in the vicinity of critical point,
the following equation can be obtained readily,
\begin{equation}
\tanh(2\kappa_c)+ \tanh\kappa_c+ \tanh(2\kappa_c) (2 b - a) + \tanh\kappa_c (b - 2 a)=1.
\end{equation}
Solution of this equation produces the value of  $0.48$ for $\kappa_c$.
The improvement of this estimation can be possible with approximating the three site correlation function
as $<\sigma><\sigma_2\sigma_3>$. Inserting this approximate relation of the three site correlation into Eq. (25), it leads to
\begin{eqnarray}
&&\tanh(2\kappa_{c})+ \tanh\kappa_{c}+ \tanh(2\kappa_{c}) (2 b - a) + \tanh\kappa_{c} (b - 2 a) {}\nonumber\\&&
+(-a\tanh(2\kappa_{c}) + b\tanh\kappa_{c})<\sigma_{2}\sigma_{3}>|_{\kappa=\kappa_{c}}=1.
\end{eqnarray}
Unfortunately, the exact value of $<\kappa_2\sigma_3>|_{\kappa=\kappa_c}$ is not obtained for honeycomb lattice in literature.
But, I think the exact values obtained for square lattice can be a relevant approximation for the term $<\sigma_2\sigma_3>|_{\kappa=\kappa_c}$. We have the value  $0.63$ for square lattice. Substituting this approximate value into the last relation
produces $\kappa_{c}=0.595$. If the exact value of critical coupling strength $\kappa_{c}=0.6585$ is considered, one can easily claim
that systematic and tractable improvement of the estimation of critical coupling strength is also possible for the honeycomb lattice
in our decimation transformation approach. 

As a next step, getting interrelation between correlation functions from the Eq. (22) can be relevant. Indeed, doing the same type of calculation as have done up to now, 
after some algebra, 
one can obtain the following result
\begin{eqnarray}
<\!\!\sigma_{0,i}\sigma_{1,i}\!\!>=\!\frac{\cosh2\kappa\tanh\kappa}{-2+4\cosh2\kappa}+\frac{1}{6}[1\!+\!3\!\tanh2\kappa\!+\!\frac{\sinh2\kappa}{-1+2\cosh2\kappa}]\!\!<\!\!\sigma_{1,i}\sigma_{2,i}\!\!>.
\end{eqnarray}
We are going to utilize this relation to obtain one of the important constants of the correlation function. In scaling theory the two site correlation function 
for $\kappa<\kappa_{c}$  is expressed as,
\begin{equation}
<\sigma_{i}\sigma_{j}>=\frac{A}{|r_{j}-r_{i}|^{\eta}}e^{-\frac{|r_{j}-r_{i}|}{\xi(\kappa)}}.
\end{equation}
Here, the critical exponent $\eta=\frac{1}{4}$ is for the 2D Ising lattices and $|r_{j}-r_{i}|$ is the distance between the $j^{th}$ and $i^{th}$ sites. 
$\xi (\kappa)$ is the correlation length which diverges (or goes to infinity) at the critical point $\kappa_{c}$. Adapting the final equation notations to the Eq.(29), it can be written at the critical point $\kappa_{c}=0.6585$ of honeycomb lattice as
\begin{equation}
A=0.1924+0.6959\frac{A}{(3)^{1/8}}.
\end{equation}
Here we used the nearest neighbor distance $a=1$, and next nearest neighbor distance equal to $\sqrt{3}$. Solution of this equation produces $A=0.4892$. It is important to notice that we have just used an exact calculation to obtain the value of $A$  for honeycomb lattice.  It is important to notice that it is easy to obtain the correlation length $\xi(\kappa)$ numerically from Eq. (29). This means that the two point correlation function can be calculated approximately in the realm of scaling theory for $\kappa<\kappa_c$.  It is also important to notice that very general two-site correlation functions have been calculated previously for square lattice \cite{McCoy2,Ghosh,Ghosh1,McCoy3}, and the value of $A$ for square lattice was obtained as $A=0.645$  by McCoy. 

\section{Conclusion and Discussion} 
In this paper, the order parameters and the two site correlations of relations of the Ising lattices are obtained straigth forwardly with the decimation transformation procedures in the presence of external field.  We have made use of these relation to study the
2D square and honeycomb Ising lattices, with particular emphasis on 
the average magnetizations and the two site correlation functions of the honeycomb and the square lattices.  The exact
expressions for three site correlation functions of square and
honeycomb lattices, obeying the relation
$<\!\!\sigma_1\sigma_2\sigma_3\!\!>=f(\kappa)<\!\!\sigma\!\!>$, have been derived
analytically. This relation indicates that the three site
correlation functions of the square and honeycomb lattice behave
like order parameters as well as the average magnetization. The obtained relations between the average magnetization and the three site correlation functions of the honeycomb and square lattices may have other important results which are not  known up to now.  I think, however,
that it is always relevant and significant  to have exact expressions for a
complex system in general.  For example, it might be a relevant testing ground
for numerical studies.

 From the two site correlation function relation of te square lattice, the four site correlation function for
square lattice are obtained. Since exact expressions for two site correlation
function of the square lattice are available from previous studies,
this can be considered as an exact expression. Besides these exact
results, mean field treatments of these lattices are taken into
account with a certain extent. As a first approximation, the order
parameter relation obtained in this paper reproduces the
previously obtained result, namely $\kappa_{c}=\frac{1}{z}$, in a
relevant manner. It is also shown that the estimation of the
critical values of coupling strength can be improved with some
traceable mean field treatments. Last, but not least, from the obtained two site correlation function of honeycomb lattice, the constant $A$ appearing in the proposed correlation function of the scaling theory has been calculated exactly for the honeycomb lattice as $A=0.4892$ in a straightforward manner. I think that this result is not only important due to its easy mathematical derivation, but also it indicates that the values of $A$ for the 2D Ising lattices are different, contrary to the assumption of the equivalence of $A$ regardless of lattice type in the scaling theory. I think that this shortcoming is the one of the most important result of this paper. 

As presented in this paper, one can use Eq. (6) and Eq. (7) derived from the decimation transformation to obtain important results which are either given in a cumbersome manner in the others considerations or are impossible to derive. We think the method derived in this paper, therefore, is relevant and fruitful in studying  Ising lattices in general, in a tractable and mathematically easy manner.

\end{document}